\documentclass[12pt]{article}
\usepackage{putex}
\usepackage{graphicx}

\begin{document}

\preprint{PUPT-2259}

\institution{PU}{Joseph Henry Laboratories, Princeton University, Princeton, NJ 08544}

\title{Gluon energy loss in the gauge-string duality}

\authors{Steven S. Gubser, Daniel R. Gulotta, Silviu S. Pufu, and F\'abio D. Rocha}

\abstract{We estimate the stopping length of an energetic gluon in a thermal plasma of strongly coupled ${\cal N}=4$ super-Yang-Mills theory by representing the gluon as a doubled string rising up out of the horizon.}

\date{March 2008}

\maketitle

\tableofcontents

\section{Introduction and summary}

Following \cite{Liu:2006ug,Herzog:2006gh,Casalderrey-Solana:2006rq,Gubser:2006bz}, considerable effort has been devoted to understanding what string theory might have to say about the physics of hard probes in heavy-ion collisions.  The key question is how quickly a hard parton loses energy as it passes through the quark-gluon plasma (QGP).  The approach of \cite{Liu:2006ug} (see also \cite{Liu:2006he}) is tied to the BDMPS jet-quenching formalism \cite{Baier:1996sk,Zakharov:1996fv,Zakharov:1997uu} (see also \cite{Baier:1998yf,Wiedemann:2000za}, and \cite{CasalderreySolana:2007pr} for a recent review), with a definition of $\hat{q}$ in terms of a partially lightlike Wilson loop.  The approaches of \cite{Herzog:2006gh,Casalderrey-Solana:2006rq,Gubser:2006bz} are limited to heavy quarks and focus on drag and stochastic forces.  Here we would like to propose an extension of the approach of \cite{Herzog:2006gh,Gubser:2006bz} to accommodate gluons in ${\cal N}=4$ super-Yang-Mills theory (SYM).  An energetic, off-shell gluon in the thermal medium should be represented as a doubled string coming up out of the horizon of the $AdS_5$-Schwarzschild geometry.  Such a string must eventually fall back into the horizon.  So the question is how far the string gets before it does so.  We will estimate the maximum penetration length $\Delta x$, as a function of the initial energy $E$, which is assumed to be much greater than the temperature.  If we define
 \eqn{HattedDefs}{
  \hat{x} = \pi T x \qquad \hat{E} = {1 \over \sqrt{g_{YM}^2 N}}
    {E \over T} \,,
 }
where $T$ is the temperature and $g_{YM}^2 N$ is the 't~Hooft coupling of SYM, then the relation we find, for sufficiently large $\hat{E}$, is
 \eqn{ApproxRelation}{
  \Delta\hat{x} \approx 0.95 {\hat E}^{1/3} \,.
 }
This relation is obtained from averaging the leading behavior of the analytic estimates \eqref{LargeESpacetime} and \eqref{LargeEWorldsheet}.  The scaling $\Delta x \propto E^{1/3}$ is different from the BDMPS scaling $\Delta x \propto E^{1/2}$, but not very different.  In order to compare with BDMPS, we make a rough operational definition of $\hat{q}$ in terms of the stopping length $\Delta x$ of a gluon of energy $E$:
 \eqn{qhatOperational}{
  \hat{q}_{\rm rough} \equiv {4E \over 3 \alpha_s (\Delta x)^2} \,.
 }
Comparing with BDMPS energy loss is hazardous because the underlying physical picture is significantly different.  Nevertheless, we plug our estimates of $\Delta x$ into \eno{qhatOperational} to obtain estimates for the corresponding value of $\hat{q}$.  To extract numerical values, we consider QCD at a temperature of $280\,{\rm MeV}$, which is representative of central gold-gold collisions at $\sqrt{s_{NN}} = 200\,{\rm GeV}$.  We exhibit our estimates in figures~\ref{ObviousPlot} and~\ref{AlternativePlot}.  These figures differ only in how we compare ${\cal N}=4$ SYM to QCD: in the nomenclature of \cite{Gubser:2006qh}, figure~\ref{ObviousPlot} uses the ``obvious'' scheme and yields $\hat{q} \approx 92 \, {\rm GeV}^2/{\rm fm}$ in the range $E = 5-25\,{\rm GeV}$ for the gluon; and figure~\ref{AlternativePlot} uses the ``alternative'' scheme and yields $\hat{q} \approx 21\,{\rm GeV}^2/{\rm fm}$ in the same energy range.  For reasons explained in \cite{Gubser:2006qh}, we prefer the alternative scheme, where comparisons are made at fixed energy density and the coupling is chosen to make the quark-anti-quark potential in ${\cal N}=4$ SYM agree as well as it can with lattice results for QCD at separations on the order of $0.25\,{\rm fm}$.  In any case, it seems clear that energy loss and thermalization as estimated from our falling string picture is more rapid than in the BDMPS formalism with $\hat{q}$ taken either from perturbative estimates or from \cite{Liu:2006ug}.  On the other hand, according to \cite{Adare:2008cg}, comparison of parton quenching model calculations \cite{Dainese:2004te,Loizides:2006cs} to PHENIX data leads to the following $3\sigma$ range for the averaged value $\langle \hat{q} \rangle$:
 \eqn{PHENIXqhat}{
  7\,{{\rm GeV^2} \over {\rm fm}} \lsim \langle \hat{q} \rangle \lsim
    28 {\rm GeV^2 \over {\rm fm}} \,,
 }
with lowest $\chi^2$ at $\langle \hat{q} \rangle \approx 13\,{\rm GeV}^2/{\rm fm}$.  It is pleasant that our estimate of $\hat{q}$ using the ``alternative'' scheme falls well inside the experimentally favored range \eno{PHENIXqhat}.  However, we emphasize that there are significant caveats to this comparison, to be discussed further in section~\ref{COMPARISONS}.

The idea that an off-shell gluon in a thermal medium should be represented as we have suggested has several antecedents, including \cite{Bak:2007fk,Alday:2007hr}.\footnote{Strings falling into anti-de Sitter space have been considered in other contexts related to heavy-ion physics in \cite{Herzog:2006gh,Lin:2006rf,Lin:2007fa}.}  In \cite{Bak:2007fk} it was argued that in computing Wilson loops at finite temperature, the configuration of two anti-parallel strings rising from the horizon to the boundary, and representing a widely separated quark-anti-quark pair, receives a color factor of $N^2$ because of the two string ends on the boundary.\footnote{When the strings are separated, they only interact by virtual exchange of closed string states, and such exchanges are suppressed by powers of $N$.  The main point of the counting powers of $N$ in \cite{Bak:2007fk} was that the $N^2$ from the horizon cancels against a $1/N^2$ suppression of this type in order to produce a final amplitude which is parametrically comparable to the one coming from a string that joins the two quarks without passing into the horizon.}  This makes sense because the $N$ D3-branes are in some sense ``behind'' the horizon, and a string ending on one of $N$ D3-branes indeed acquires a fundamental or anti-fundamental charge.  We had been in the habit of thinking of color charges living on the boundary of $AdS_5$-Schwarzschild, but it seems more faithful to the D-brane origin of a near-extremal black 3-brane for the color factor to come from ends at the horizon.  Seen in this light, the trailing string of \cite{Herzog:2006gh,Gubser:2006bz} derives its fundamental color charge from the fact that it actually passes through the horizon.  (See \cite{Casalderrey-Solana:2007qw} for a particularly clear exposition of the geometry of the trailing string.)  What could be more natural, then, than to turn the heavy quark into an energetic gluon by letting the string double over on itself and pass back down into the horizon, rather than rising all the way up to the conformal boundary?

The work of \cite{Alday:2007hr} employs a zero-temperature limit of approximately this construction to consider collisions of gluons.  In pure $AdS_5$, however, one can insist upon the view that color degrees of freedom ``live'' on the boundary: upon conformal compactification to global $AdS_5$, the endpoints of the strings are seen to rise back up to the boundary at a point which is infinitely far from the collision region.  In \cite{Ito:2007zy}, some results of \cite{Alday:2007hr} were extended to finite temperature.  Whereas in \cite{Alday:2007hr,Ito:2007zy} the focus was on scattering amplitudes of several hard gluons, here we are interested in the propagation of a single hard gluon through the thermal medium.

The organization of the rest of this paper is as follows.  In section~\ref{ENERGY} we explain how to calculate the energy of a gluon represented as a doubled string rising vertically up from the horizon.  Section~\ref{TRAILING} shows how to carry out an analogous computation when the shape of the string is part of the trailing string.  Section~\ref{GEODESICS} detours into the computation of lightlike geodesics in the $AdS_5$-Schwarzschild geometry.  Section~\ref{SPACETIME} presents estimates of $\Delta\hat{x}$ as a function of $\hat{E}$ using the lightlike geodesics discussed in section~\ref{GEODESICS}.  Section~\ref{WORLDSHEET} presents estimates of $\Delta\hat{x}$ as a function of $\hat{E}$ using lightlike geodesics on the worldsheet of the trailing string.  Section~\ref{COMPARISONS} describes comparisons with the BDMPS energy-loss formalism, expanding on the brief discussion above.  Section~\ref{NULL} describes a lightlike limit of the falling string which is analytically tractable.  We end in section~\ref{DISCUSSION} with some discussion of possible extensions of the falling string picture.

\section{Estimating the energy of a doubled string}
\label{ENERGY}

\begin{figure}
  \centerline{\includegraphics[width=6in]{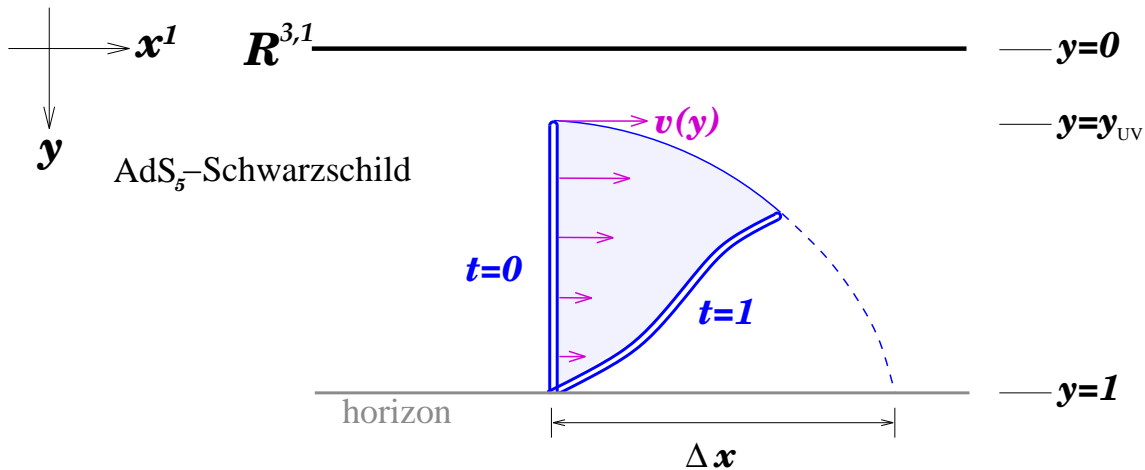}}
  \caption{If a string starts at $t=0$ in a straight up-and-down configuration, it doesn't hold its shape as time evolves forward.  At a later time, indicated as $t=1$ in the figure, one must solve difficult classical equations of motion to find the shape of the string.  But because of the infinite redshift characteristic of black hole horizons, the point where the string comes out of the horizon cannot move at all.  At times $t>1$, the string continues to fall down toward the horizon.  Although it takes an infinite time to fall all the way in, it only propagates a finite distance $\Delta x$ forward.}\label{VerticalString}
 \end{figure}

If we accept that an off-shell gluon traveling through a thermal medium should be represented as a string with both its endpoints passing through the horizon of $AdS_5$-Schwarzschild, the next question is what the shape of the string should be.  A natural first guess is that it should be straight up and down.  The problem is that a string that is straight up and down at $t=0$ will not hold its shape as it moves in the positive $x^1$ direction.  This is illustrated in figure~\ref{VerticalString}.  To demonstrate that the string can't stay vertical, recall first the $AdS_5$-Schwarzschild metric:
 \eqn{AdSSchwarzschild}{
  ds^2 = G_{\mu\nu} dx^\mu dx^\nu =
   {L^2 \over z^2} \left( -h dt^2 + d\vec{x}^2 +
    {dz^2 \over h} \right) \qquad\hbox{where}\qquad
    h = 1-{z^4 \over z_H^4} \,,
 }
and the Nambu-Goto action:
 \eqn{SNG}{
  S_{NG} = -{1 \over 2\pi\alpha'} \int d^2 \sigma \sqrt{-g}
   \qquad\hbox{where}\qquad
  g_{\alpha\beta} = G_{\mu\nu} \partial_\alpha X^\mu
    \partial_\beta X^\nu \,.
 }
Our convention is to use indices $\alpha,\beta$ for the string worldsheet coordinates $\sigma^\alpha$, and capital $X^\mu = X^\mu(\sigma^\alpha)$ for the embedding coordinates of the classical solution under consideration.  The worldsheet current of spacetime stress-energy is
 \eqn{Palphamu}{
  P^\alpha_\mu = -{1 \over 2\pi\alpha'} g^{\alpha\beta}
    G_{\mu\nu} \partial_\beta X^\mu \,,
 }
and the equations of motion following from \eno{SNG} are $\nabla_\alpha P^\alpha_\mu = 0$.  Here $\nabla_\alpha$ is the covariant derivative with respect to the worldsheet metric $g_{\alpha\beta}$.  Five-dimensional indices like $\mu$ are treated as scalars with respect to $\nabla_\alpha$.  Suppose we start the string in a straight up-and-down configuration at time $t=0$, as illustrated in figure~\ref{VerticalString}, with an initial velocity profile $v=v(z)$ in the $x^1$ direction.
Using coordinates $\sigma^\alpha = (t,z)$, this means that, at $t=0$,
 \eqn{FoundPalphamu}{
  {dX^\mu \over d\sigma^\alpha} =
   \begin{pmatrix} 1 & 0 \\ v & 0 \\ 0 & 0 \\ 0 & 0 \\ 0 & 1
    \end{pmatrix} \qquad
  g_{\alpha\beta} = {L^2 \over z^2} \begin{pmatrix}
   -h+v^2 & 0 \\ 0 & 1/h \end{pmatrix} \qquad
  P^\alpha_\mu = {1 \over 2\pi\alpha'} \begin{pmatrix}
    -{h \over h-v^2} & {v \over h-v^2} & 0 & 0 & 0 \\
    0 & 0 & 0 & 0 & 1
   \end{pmatrix}
 }
Already from \eqref{FoundPalphamu} we can see that there is a depth-dependent limit on the velocity: $v < \sqrt{h}$ in order for the metric to have Lorentzian signature.  Variants of this speed limit have been discussed in a number of papers, including \cite{Herzog:2006gh,Peeters:2006iu,Liu:2006nn,Chernicoff:2006hi,Ejaz:2007hg,Argyres:2008eg}.  It implies that the string indeed cannot stay upright: it would do so only if $v$ is constant in $z$, and the only way that can be true is if $v=0$.  To put it another way, the point where the string comes out of the horizon cannot move in the $x^1$ direction.  All the string can do is to fall over into the horizon.

Assuming the speed limit $v < \sqrt{h}$ is satisfied everywhere along the string, the total momentum of the string can be computed as
 \eqn{TotalPmu}{
  p_\mu = \int dz \, \sqrt{-g} P_\mu^t =
    {L^2 \over 2\pi\alpha'} \int {dz \over z^2} \,
     \sqrt{h \over h-v^2} \begin{pmatrix} -1 &
     {v \over h} & 0 & 0 & 0 \end{pmatrix} \,.
 }
The first four components $p_m$ of $p_\mu$ can be identified with the four-momentum of the boundary gauge theory.  This is because they are momenta defined in reference to Killing vectors $\partial/\partial x^m$ for $m=0,1,2,3$.  (We use mostly minus signature, so the energy is $-p_0>0$.)  The fifth component $p_z$ does not have such a simple interpretation.  The integral in \eno{TotalPmu} should be taken over the intersection of the string worldsheet with the hypersurface $t=0$.  More specifically, it should be taken over that part of the string worldsheet that is outside the horizon.  Because this part is doubled over and rises to a minimum depth $z_{\rm UV}$, we find
 \eqn{zUVint}{
  p_\mu = {L^2 \over \pi\alpha'} \int_{z_{\rm UV}}^{z_H}
   {dz \over z^2} \, \sqrt{h \over h-v^2} \begin{pmatrix} -1 &
     {v \over h} & 0 & 0 & 0 \end{pmatrix} \,.
 }
There are two subtleties that affect \eno{zUVint}:
 \begin{itemize}
  \item The kink at $z=z_{\rm UV}$ can support a finite lightlike momentum $\delta p_\mu$, which would have to be added to $p_\mu$.  Light-like means $G^{\mu\nu} \delta p_\mu \delta p_\nu = 0$, which is to say $h (\delta p_0)^2 = (\delta p_1)^2$ if $\delta p_\mu=0$ for $\mu>1$.
  \item One could choose to run the integral over the part of the worldsheet behind the horizon.  Not doing so is a physical choice, motivated by the fact that nothing behind the horizon can classically influence what's outside.  We regard whatever the string does behind the horizon as part of the dynamics of the thermal medium.
 \end{itemize}
The result \eno{zUVint} is analogous to the expressions $E = m\gamma$ and $p = mv\gamma$ for a massive particle.  It shows that there are qualitatively different ways in which to make the string representing the gluon highly energetic: one may either take $z_{\rm UV} \to 0$, or make the local ``Lorentz'' factor $1/\sqrt{h-v^2}$ big over some portion of the string worldsheet.

It has recently been emphasized in \cite{LindenLevy:2007gd} that a quasi-particle description of the QGP may not be valid.  A quasi-particle picture is even less likely to capture the physics of strongly coupled ${\cal N}=4$ super-Yang-Mills, where the weakly coupled degrees of freedom are manifest only in the dual gravitational description.  On the other hand, as long as a gluon has energy and momentum much greater than the temperature, it scarcely notices the thermal bath, and there should be an approximately unique way to describe it.  The small $z$ region of $AdS_5$-Schwarzschild is where the presence of the horizon doesn't matter, and it is associated with UV physics because an object there translates into a tightly localized or highly energetic configuration in gauge theory.  So we recover the intuition that a hard gluon can be defined with little ambiguity by assuming that most of its momentum comes from the small $z$ region:
 \eqn{pUV}{
  p_\mu = p_\mu^{\rm UV} + \hbox{(infrared effects)} \qquad\hbox{where}
   \qquad p_\mu^{\rm UV} = {L^2 \over \pi\alpha'}
   {1 \over \sqrt{1-v^2}}
   {1 \over z_{\rm UV}} \begin{pmatrix} -1 & v & 0 & 0 & 0 \end{pmatrix}
     \,.
 }
The expression $p_\mu^{\rm UV}$ comes from setting $h=1$ and $z_H=\infty$ in \eno{zUVint} before carrying out the $z$ integration: that is, we ignore the bath altogether.  Depending on the context, a better approximation may be needed: for instance, one might want to replace $1-v^2$ by $h_{\rm UV}-v^2$ so that the limit on $v$ previously discussed is correctly implemented for finite but small $z_{\rm UV}$.

The result \eno{pUV} also focuses attention on the fact that the string describes an off-shell object:
 \eqn{QsquaredDef}{
  Q^2 \equiv -(p_m^{\rm UV})^2 = \left( {L^2 \over \pi\alpha'
    z_{\rm UV}} \right)^2 > 0 \,.
 }
Recall that we use mostly plus signature: thus with the explicit sign included in \eno{QsquaredDef}, $Q^2>0$ means timelike momentum.  We do not know how to represent a gluon with $Q^2<0$.  The limit $v \to 1$ from below with $p_0$ held fixed corresponds to taking the gluon on-shell.  But this limit is not available for $T>0$, because eventually it would force $z_{\rm UV}$ to become greater than $z_H$, and ignoring the bath would then be wrong.  This discussion highlights a crucial difference between our approach and the more conventional BDMPS treatment, where the first step is to make an eikonal approximation where the gluon travels strictly at the speed of light.  While such an approximation is reasonable in perturbation theory and makes sense for sufficiently energetic probes of a finite-sized medium in an asymptotically free theory like QCD, we harbor some doubts about the consistency of expanding around light-like trajectories of charged particles in theories such as ${\cal N}=4$ super-Yang-Mills where the coupling is finite even in the ultraviolet.  In any case, a particle which propagates only a finite distance through the medium is not on-shell because it is not an asymptotic state.  Thus we are more reassured than concerned over being forced to take $v<1$.

\section{Energetic gluons and the trailing string}
\label{TRAILING}

With an approximate expression \eno{pUV} in hand for the momentum $p_\mu$ of a hard gluon represented by a doubled string, the next thing  we should ask is how far in the $x^1$ direction the doubled string travels before it falls into the black hole.  Falling into the black hole corresponds to thermalization in the dual gauge theory.  So if the string travels a distance $\Delta x$ before falling in completely, we expect that $\Delta x$ should be identified, at least approximately, as the stopping distance of the hard gluon in the medium.  Both the energy, $-p_0$, and the penetration length, $\Delta x$, are functions of $v$ and $z_{\rm UV}$, so the question of maximum penetration length can be phrased as maximizing $\Delta x$ with $-p_0$ held fixed.  This extremization problem is challenging because evaluating $\Delta x$ involves solving the non-linear (but classical) equations of motion for the string, starting from an initial state which is only approximately specified.  Instead of tackling this problem head-on, let's go back to the observation that the straight up-and-down string configuration considered in section~\ref{ENERGY} doesn't hold its shape for $t>0$.  Is there is some other string configuration which does?  The trailing string of \cite{Herzog:2006gh,Gubser:2006bz} suggests itself immediately: it is a steady state solution of a string whose endpoint on the conformal boundary is required to move with a definite velocity $v$.  That shape is specified in the gauge $\sigma^\alpha = (t,z)$ by the embedding
 \eqn{Xembed}{
  X^1 = v \left[ t + \xi(z) \right] \qquad\hbox{where}\qquad
   \xi(z) = -{z_H \over 4i} \left(
    \log {1-iy \over 1+iy} + i \log {1+y \over 1-y} \right) \,.
 }
Here we have introduced a rescaled depth coordinate
 \eqn{yDef}{
  y = {z \over z_H} \,.
 }
We will persist in using worldsheet coordinates $\sigma^\alpha = (t,z)$, but we express results, such as the right hand side of \eno{Xembed}, in terms of $y$ when convenient.

\begin{figure}
  \centerline{\includegraphics[width=6in]{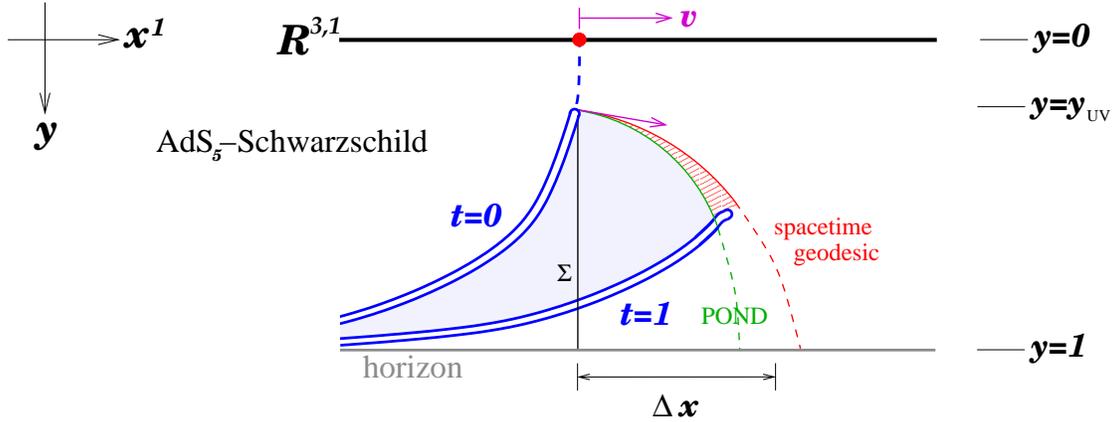}}
  \caption{A string starts at $t=0$ in the shape of a falling string that extends up to a finite minimum depth, $y=y_{\rm UV}$.  (Recall that $y=z/z_H$.)  If the string extended up to the boundary, as shown with a dashed curve, its endpoint would move at a speed $v$.  The lightly shaded region, below the trajectory labeled POND, retains the shape of the trailing string.  At times $t>0$ (for instance at the time denoted $t=1$ in the figure), the string probably projects somewhat beyond the POND trajectory into the narrow region between it and the null spacetime geodesic.  The POND trajectory and the spacetime geodesic are mutually tangent at the point $t=0$, $y=y_{\rm UV}$.}\label{FallingString}
 \end{figure}

Studies in \cite{Herzog:2006gh} (see also \cite{Gubser:2006nz}) show that the trailing string is stable against small perturbations.  As an alternative initial condition to the straight up-and-down string discussed previously, let's therefore consider a string which starts at $t=0$ in the shape \eno{Xembed}, except that it rises to a finite minimum depth $z_{\rm UV}$ before doubling back over itself and going back down into the horizon.  See figure~\ref{FallingString}.
What we will find is that although this string doesn't hold its shape exactly, it ``almost'' does, in a sense we will describe in section~\ref{WORLDSHEET}, when its momentum is large.

The worldsheet current of spacetime energy-momentum on the trailing is computed as follows:
 \eqn[c]{TrailingEnergy}{
  {dX^\mu \over d\sigma^\alpha} =
    \begin{pmatrix} 1 & 0 \\ v & -vy^2/h \\ 0 & 0 \\ 0 & 0 \\ 0 & 1
     \end{pmatrix} \qquad
  g_{\alpha\beta} = {L^2 \over z_H^2 y^2} \begin{pmatrix}
    -h+v^2 & -v^2 y^2/h \\ -v^2 y^2/h &
     (h+v^2-hv^2)/h^2 \end{pmatrix}  \cr
  P^\alpha_\mu = {1 \over 2\pi\alpha' h (1-v^2)}
   \begin{pmatrix} -h-v^2+hv^2 & v & 0 & 0 & {v^2 y^2 \over h} \\
    -h v^2 y^2 & hv y^2 & 0 & 0 & -h + v^2 \end{pmatrix} \,.
 }
The total momentum of the doubled string may be computed at $t=0$ in a fashion analogous to \eno{TotalPmu}--\eno{zUVint}:
 \eqn{TrailingPmu}{
  p_\mu = \int dz \, \sqrt{-g} P^t_\mu =
   {L^2 \over \pi\alpha' z_H} {1 \over \sqrt{1-v^2}} \int_{y_{\rm UV}}^1
    {dy \over hy^2} \begin{pmatrix} -h-v^2+hv^2 & v & 0 & 0 &
      {v^2 y^2 \over h} \end{pmatrix} \,.
 }
It is important in \eno{TrailingPmu} that we stipulated that $t=0$, because for $t>0$ the string will not hold its shape precisely: more on this later.

The four-dimensional components $p_m$ of the momentum have a logarithmic divergence at $y=1$.  (The last component, $p_5$, diverges as a power, $1/(1-y)$.)  In the original context of a heavy quark propagating through the thermal medium \cite{Herzog:2006gh}, this infrared divergence owes to the fact that the trailing string is the shape attained in a late-time limit, and it accounts for the large amount of energy that has already been transferred (or mostly transferred) from the quark to the thermal bath.\footnote{The infrared divergence provides an extreme example of how various shapes of the string describe various states of the gluon and the bath.  The infrared tail of the trailing string encodes the interaction of the gluon with the medium in such a way as to form a sonic boom plus a diffusion wake.}  In the current context, the divergence should be regulated somehow, because we have in mind creating a gluon with large but finite energy at $t=0$ and then asking how far it propagates.  The simplest regulator is simply to cut off the integral in \eno{TrailingPmu} at some $y_{\rm IR}$ slightly less than $1$.

Instead of measuring the energy and momentum of the string at $t=0$, one can obtain a divergence-free definition of $p_m$ by calculating the amount of energy and momentum in the string that makes it past a fixed value of $x^1$.  More precisely, we should compute the flux of worldsheet energy-momentum $P_\mu^\alpha$ through the intersection of the string worldsheet with the hypersurface $x^1 = {\rm constant}$.  To prepare for this computation, consider a general conserved worldsheet current $Q^\alpha$.  Its flux through a curve $\Sigma$ on the worldsheet, specified by $\sigma^\alpha = \sigma^\alpha(\eta)$, is
 \eqn{WSFlux}{
  Q = \int_{\Sigma} d \eta \, \sqrt{-g} \, \epsilon_{\alpha \beta} Q^\alpha {d\sigma^\beta \over d \eta} \,,
 }
where $\epsilon_{\alpha\beta}$ is the antisymmetric tensor normalized so that $\epsilon_{12} = 1$.  Now let $\Sigma$ be the curve at $x^1=0$ on the trailing string, as shown in figure~\ref{FallingString}.  Using $\eta=z$ as a parameter for this curve, and replacing $Q^\alpha$ by $P^\alpha_\mu$ in \eno{WSFlux}, one obtains
 \eqn{Fixedx1pmu}{
  p_\mu^{\textrm{fixed $x^1$}} &= \int_{z_{\rm UV}}^{z_H} dz\, \sqrt{-g} \left[P_\mu^t -
   \left({\partial t \over \partial z} \right)_{x^1} P_\mu^z \right] \cr
   &= {L^2 \over \pi\alpha' z_H} {1 \over \sqrt{1-v^2}}
    \int_{y_{\rm UV}}^1 {dy \over y^2} \begin{pmatrix} -1 & v & 0 & 0 &
      {y^2 \over h} \end{pmatrix}\,,
 }
where in the second expression the derivative $\left({\partial t \over \partial z} \right)_{x^1}$ is taken at constant $x^1$.  Unlike in \eqref{TrailingPmu}, the integrals defining the four-dimensional components $p_m^{\textrm{fixed $x^1$}}$ are convergent, so they don't require an IR cutoff.

We are reassured to observe that for $y_{\rm UV} \ll 1$, we recover the form \eno{pUV} from both \eqref{TrailingPmu} and \eqref{Fixedx1pmu}.  We will ignore the possibility of an additional lightlike contribution to $p_\mu$ from the tip of the string, where it doubles over.

\section{Lightlike geodesics in $AdS_5$-Schwarzschild}
\label{GEODESICS}

The tip of an open string, or a doubled string, must move at the speed of light.  But it usually does not follow a lightlike geodesic, because it is being pulled in some direction by the rest of the string.  In the case under consideration, the pull is downward (in the positive $z$ direction) and backward (in the negative $x^1$ direction).  So to find an upper bound on how far the string gets in the positive $x^1$ direction before falling through the horizon, we could consider the trajectory of a lightlike particle that starts at the tip of the string at $t=0$ and falls into the horizon without experiencing the pull of the string.  To this end, let's work out free particle trajectories in the $AdS_5$-Schwarzschild geometry.  Parameterizing the particle's worldline $X^\mu = X^\mu(\eta)$ with an arbitrary variable $\eta$, the action may be expressed as
 \eqn{AffineAction}{
  S = {1 \over 2} \int d\eta \, \left[
    {1 \over e} G_{\mu\nu} {dX^\mu \over d\eta}
     {dX^\nu \over d\eta} - m^2 e \right] \,.
 }
Here $m$ is the mass (eventually to be taken to $0$) and $e$ is a Lagrange multiplier.  For $m\neq 0$, one may use the constraint equation for $e$ to eliminate $e$ from the action.  After doing so, the action \eno{AffineAction} reduces to the standard one,
 \eqn{StandardAction}{
  S = -\int ds \, m \,.
 }
An advantage of \eno{AffineAction} is that its $m \to 0$ limit correctly describes the dynamics of massless particles.

Let's work in a gauge where $\eta = z$ and consider trajectories of the form
 \eqn{OneDirection}{
  X^0 = X^0(z) \qquad X^1 = X^1(z) \qquad X^2 = X^3 = 0 \,.
 }
Then
 \eqn{Sstatic}{
  S = \int dz \, {\cal L} \qquad\hbox{where}\quad
   {\cal L} = {L^2 \over 2ez^2} \left( -h (X^{0\prime})^2 +
     (X^{1\prime})^2 + {1 \over h} \right) -
     {1 \over 2} m^2 e \,,
 }
where primes indicate $d/dz$.  One may immediately form conserved momenta
 \eqn{Momenta}{
  p_0 \equiv {\partial {\cal L} \over \partial X^{0\prime}} =
    -{L^2 \over ez^2} h X^{0\prime} \qquad
  p_1 \equiv {\partial {\cal L} \over \partial X^{1\prime}} =
    {L^2 \over ez^2} X^{1\prime} \,.
 }
The equation of motion for $e$ is an algebraic constraint:
 \eqn{eConstraint}{
  e = \pm {L^2 \over z^2} {1 \over \sqrt{p_0^2 - h p_1^2 -
    h m^2 L^2/z^2}} \,.
 }
In order to make energy positive, $p_0$ should be negative, so we should choose the plus sign in \eno{eConstraint} for trajectories (or segments of trajectories) where $z$ increases as $t$ increases, and the minus sign for trajectories where $z$ decreases as $t$ increases.  Hereafter we will always choose the plus sign, corresponding to particles falling down toward the horizon.

\begin{figure}
  \centerline{\includegraphics[width=3.5in]{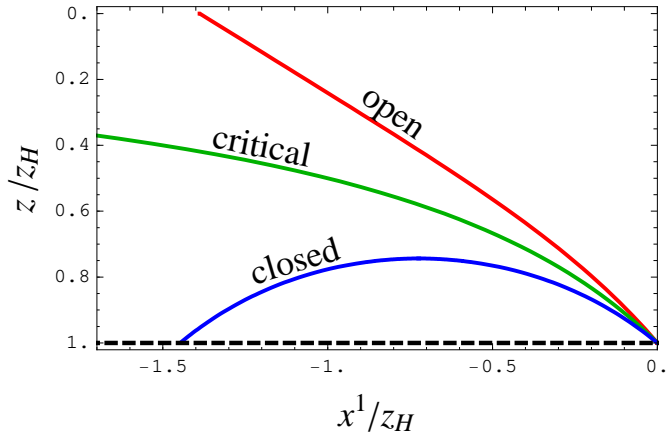}}
  \caption{Typical orbits for a massless particle in $AdS_5$-Schwarzschild, all leading into the horizon (the dashed black line) at $x^1=0$.  An open orbit is shown in red; the critical orbit is shown in green; and a closed orbit is shown in blue.}\label{MasslessOrbits}
 \end{figure}

The particle trajectories $X^\mu = X^\mu(z)$ can be determined using \eno{Momenta}--\eno{eConstraint} once one specifies $p_0$ and $p_1$.  In the massless limit, the shape of the orbits $X^1 = X^1(z)$ is determined from
 \eqn{dzdx}{
  {dX^1 \over dz} = X^{1\prime} = {ez^2 \over L^2} p_1 =
    -{p_1/p_0 \over \sqrt{1 - h p_1^2 / p_0^2}} \,.
 }
To perform the integral in \eno{dzdx} one needs elliptic functions, and the explicit result is not very enlightening.  But for $p_1=-p_0$, the result is very simple:
 \eqn{SpecialOrbit}{
  {X^1 \over z_H} = K - {z_H \over z} = K - {1 \over y} \,,
 }
where $K$ is a constant of integration.  One may also straightforwardly show that
 \eqn{SpecialTime}{
  X^0 = X^1 - \xi(z) \,,
 }
where $\xi(z)$ is as defined in \eno{Xembed}.  We will refer to \eno{SpecialOrbit} as the critical orbit, because for $p_1<-p_0$ the orbits intersect both the conformal boundary and the horizon, while for $p_1>-p_0$ the orbits begin and end at the horizon: see figure~\ref{MasslessOrbits}.  In all cases, getting to or from the horizon takes an infinite amount of coordinate time $X^0$.

\section{Estimating the penetration depth using spacetime geodesics}
\label{SPACETIME}

To find an approximate upper bound on the penetration depth, we should start a massless particle at the tip of the string with a physically motivated choice of $p_\mu$.  But how should we make this choice?  To answer this question, let's have a look at the geometry of the trailing string in the hyperplane $x^2=x^3=0$.  Because the intrinsic geometry is Lorentzian, there is a basis of lightlike vectors, $\ell^\alpha$ and $k^\alpha$, on the worldsheet: using coordinates $\sigma^\alpha = (t,z)$ as before,
 \eqn{lkDef}{
  \ell^\alpha = \begin{pmatrix} {-v^2 \sqrt{1-h} + h \sqrt{1-v^2}
    \over h (h-v^2)} \\[3pt] 1 \end{pmatrix} \qquad
  k^\alpha = \begin{pmatrix} 1 \\[3pt]
   {-h (h-v^2) \over v^2 \sqrt{1-h} + h \sqrt{1-v^2}} \end{pmatrix}
    \,.
 }
All components of $\ell^\alpha$ and $k^\alpha$ are non-singular functions of $z$, and all are positive except for $k^z$, which is positive only when $h < v^2$, meaning below the worldsheet horizon at $z = z_H \sqrt[4]{1-v^2}$.  One may correspondingly form spacetime vectors $\ell^\mu = \ell^\alpha \partial X^\mu / \partial\sigma^\alpha$ and $k^\mu = k^\alpha \partial X^\mu / \partial\sigma^\alpha$.  To complete a basis for the $x^2 = x^3 = 0$ hyperplane one may add the vector $n^\mu$ normal to the worldsheet: it is defined up to an overall factor by the equations $n_\mu \partial X^\mu/\partial\sigma^\alpha=0$, or equivalently $n_\mu \ell^\mu = n_\mu k^\mu = 0$, and one easily finds
 \eqn{FoundNmu}{
  n_\mu = \begin{pmatrix} -v & 1 & 0 & 0 & {v\sqrt{1-h} \over h}
    \end{pmatrix} \,.
 }
The vector $\ell^\mu$ points in the direction of a lightlike signal traveling down the string.  A sensible initial condition for the massless particle whose trajectory is supposed to approximately bound the motion of the string worldsheet is $p_\mu \propto \ell_\mu$.  Combining this initial condition with \eno{dzdx}, we can immediately calculate how far the massless particle gets in the positive $x^1$ direction before falling into the horizon:
 \eqn[c]{FoundDeltaXone}{
  \Delta x_{\rm spacetime} = -z_H \int_{y_{\rm UV}}^1 dy {p_1/p_0 \over
    \sqrt{1 - (1-y^4) p_1^2/p_0^2}} \qquad\hbox{where}
      \cr\noalign{\vskip1\jot}
  {p_1 \over p_0} = v {\sqrt{1-v^2} - y_{\rm UV}^2 \over
    v^2 y_{\rm UV}^2 - (1-y_{\rm UV}^4) \sqrt{1-v^2}} \,.
 }
As remarked previously, the integral can be done in terms of elliptic functions, but the explicit form is unenlightening.  The subscript ``spacetime'' in \eno{FoundDeltaXone} reminds us that the calculation hinges on lightlike geodesics in the $AdS_5$-Schwarzschild spacetime.

\begin{figure}
  \centerline{\includegraphics[width=6.5in]{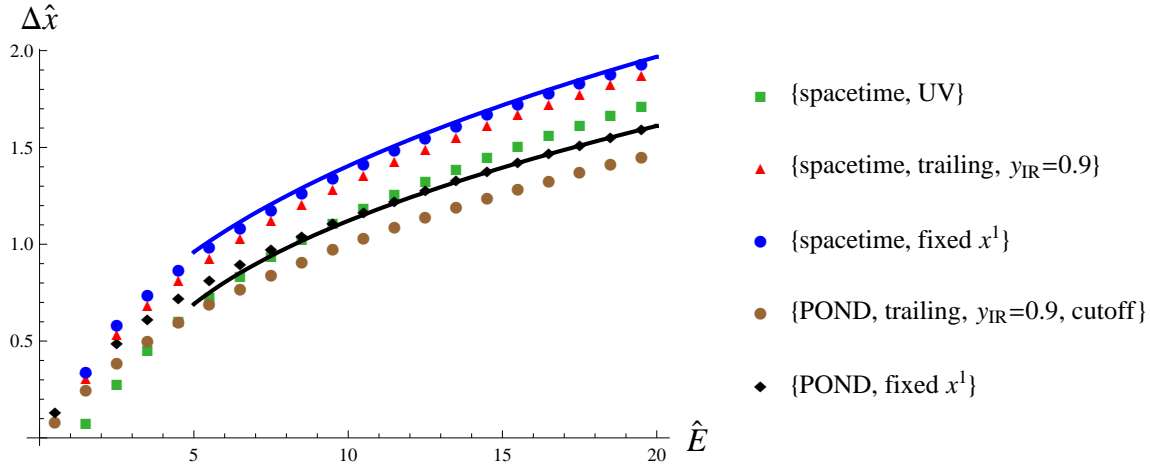}}
  \caption{Evaluations of the maximum penetration depth $\Delta \hat{x}_A(\hat{E})$ with a variety of assumptions.  The solid blue curve shows the analytic approximation \eqref{LargeEWorldsheet} to the blue circles, and the solid black curve shows the analytic approximation \eqref{LargeESpacetime} to the black diamonds.}\label{CombinedPlot}
 \end{figure}

The extremization problem that we set out to solve was to maximize the distance $\Delta x$ traveled by a classical string with fixed energy $E = -p_0$, starting from an initial configuration with both ends passing through the horizon.  What we can now do instead is to maximize $\Delta x_{\rm spacetime}$ subject to fixed energy.  We will consider three ways of defining the energy:
 \eqn{Eapprox}{
  E_{\rm UV} = {L^2 \over \pi\alpha' z_H} {1 \over \sqrt{1-v^2}}
    {1 \over y_{\rm UV}} \,,
 }
which comes from \eno{pUV};
 \eqn{Etrailing}{
  E_{\rm trailing} = {L^2 \over \pi\alpha' z_H}
    {1 \over \sqrt{1-v^2}} \int_{y_{\rm UV}}^{y_{\rm IR}}
    {dy \over hy^2} (h+v^2-hv^2) \,,
 }
which comes from \eqref{TrailingPmu}; and
 \eqn{Efixedx1}{
  E_\textrm{fixed $x^1$} = {L^2 \over \pi\alpha' z_H}
    {1 \over \sqrt{1-v^2}} \left({1\over y_{\rm UV}} - 1 \right) \,,
 }
which comes from \eqref{Fixedx1pmu}.  Evidently, the dimensionful parameters $L$, $\alpha'$, and $z_H$ enter into these expressions only as multiplicative prefactors.  So it is convenient to scale them out as in \eno{HattedDefs} by defining
 \eqn{HattedDefsAgain}{
  \hat{x}^1 = {x^1 \over z_H} = \pi T x \qquad\qquad
  \hat{E} = {\pi\alpha' z_H \over L^2} E =
    {1 \over \sqrt{g_{YM}^2 N}} {E \over T} \,,
 }
where we have used the standard relations
 \eqn{StandardRelations}{
  z_H = {1 \over \pi T} \qquad\qquad {L^2 \over \alpha'} =
    g_{YM}^2 N
 }
for ${\cal N}=4$ super-Yang-Mills theory.  The dimensionless quantities $\Delta \hat{x}_{\rm spacetime}$, $\hat{E}_{\rm UV}$, and $\hat{E}_\textrm{fixed $x^1$}$ are functions only of $v$ and $y_{\rm UV}$, and $\hat{E}_{\rm trailing}$ is a function only of $v$, $y_{\rm UV}$, and $y_{\rm IR}$.  Let's regard $y_{\rm IR}$ as a fixed cutoff.  Then the extremization of $\Delta \hat{x}_{\rm spacetime}$ with either $\hat{E}_{\rm UV}$, $\hat{E}_{\rm trailing}$, or $\hat{E}_\textrm{fixed $x^1$}$ held equal to some fixed value $\hat{E}$ is a well-defined problem, and we shall denote the result $\Delta\hat{x}_A(\hat{E})$.  The index $A$ labels the assumptions that went into the calculation.  For example, if we held $\hat{E}_{\rm UV}$ fixed in an extremization of $\Delta \hat{x}_{\rm spacetime}$, then we would say $A = \{{\rm spacetime},{\rm UV}\}$.  If instead we held $\hat{E}_{\rm trailing}$ fixed, say with $y_{\rm IR} = 0.9$, then we would say $A = \{{\rm spacetime},{\rm trailing},y_{\rm IR}\!=\!0.9\}$.  In figure~\ref{CombinedPlot} we show a number of evaluations of $\Delta\hat{x}_A(\hat{E})$ for several different choices of assumptions.

As can be seen from figure~\ref{CombinedPlot}, the functional dependence $\Delta \hat{x}_A(\hat{E})$ is similar for the various sets $A$ of assumptions.  To better understand this functional dependence, we computed a series expansion at large $\hat{E}$ for $A = \{{\rm spacetime},\textrm{fixed $x^1$}\}$:
 \eqn{LargeESpacetime}{
  \Delta \hat{x}_{{\rm spacetime},\textrm{fixed $x^1$}} = 1.0185 \hat{E}^{1/3} - 0.8180 + 0.052 \hat{E}^{-1/3} + 0.017 \hat E^{-2/3} + {\cal O}(\hat E^{-1})\,.
 }
As shown in figure~\ref{CombinedPlot}, this expression provides a fairly good approximation to the numerical evaluations.  We defer the explanation of how we derived \eqref{LargeESpacetime} until the end of section~\ref{WORLDSHEET}, where the same method will be described in a slightly simpler setting.

\begin{figure}
  \centerline{\includegraphics[width=6.5in]{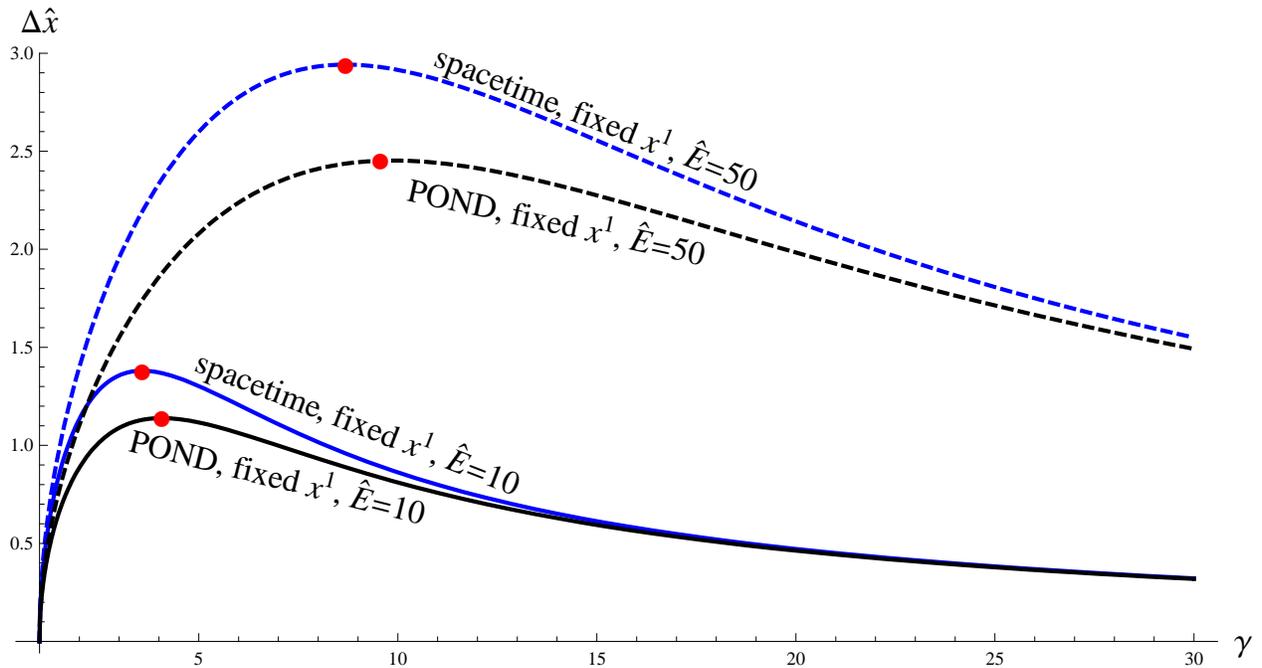}}
  \caption{The penetration length $\Delta \hat{x}$ as a function of $\gamma$ for fixed $\hat{E}$. The blue curves represent the penetration length for
  a spacetime geodesic, while the black curve represent the same quantity for the point of no disturbance (POND), which is explained in section~\ref{WORLDSHEET}. For all curves, the
  energy was computed at fixed $x^1$ using \eqref{Efixedx1}. The red points mark the maxima of each curve and correspond to data points in figure~\ref{CombinedPlot}.}\label{DeltaXProfilePlot}
 \end{figure}

To illustrate more explicitly the nature of the extremization problem, in figure~\ref{DeltaXProfilePlot} we plot $\Delta\hat{x}_{\rm spacetime}$ as a function of $\gamma=1/\sqrt{1-v^2}$ at fixed $\hat E_\textrm{fixed $x^1$}$.  We also plot $\Delta\hat{x}_{\rm POND}$, a quantity to be explained in section~\ref{WORLDSHEET}.  It is worth noting that the maximum in $\Delta\hat{x}$ is fairly broad in $\gamma$, and it gets broader as $\hat{E}$ increases.

\section{Estimating the penetration depth using worldsheet geodesics}
\label{WORLDSHEET}

As we have described, a lightlike geodesic emanating from the tip of the doubled string provides an approximate upper bound on how far its trajectory reaches forward in the $x^1$ direction---approximate because the selection of initial direction is a physical choice based on our expectations of what a typical initial state of the string should be.  There remains the possibility that the string will always fall into the horizon much more quickly than the lightlike geodesics do.  Ideally we would like to have a lower bound to show that this does not happen.  But recall the nature of the extremization problem: the maximum penetration depth $\Delta \hat{x}(\hat{E})$ is the furthest that a string can get with specified initial energy.  It's easy to see how it could get much less far: by taking both $v$ and $y_{\rm UV}$ very small, with total energy held fixed, one finds configurations that fall almost straight down into the horizon despite having a large energy.  What we really want, then, is a demonstration that there is {\it some} string configuration which gets almost as far as the lightlike geodesics we have already studied.  This turns out to be straightforward, as we will now discuss.

The configuration we want to study is again the doubled trailing string shown in figure~\ref{FallingString}.  We first mentioned it in connection with the hope that it would almost hold its shape as it falls into the horizon.  Here's an argument that it does.  First imagine the original trailing string of \cite{Herzog:2006gh,Gubser:2006bz}, which continues all the way up to the conformal boundary.  By construction, it holds its shape for all time.  Now, at time $t=0$, let's cut the string at some finite $z_{\rm UV}$, and let's keep track only of the lower part of it (i.e.~the part with $z>z_{\rm UV}$).  For some time $t$ slightly greater than $0$, most of this lower part of the string must be in the same shape that it would have been had we not cut the string, simply because it hasn't ``figured out'' that the cut occurred.  More precisely, if a lightlike signal hasn't had time to propagate down the string below a certain depth $z_*$, then the part of the string with $z>z_*$ must be in the same configuration that it would have been had we not cut the string.  The motion of the part of the trailing string which the lightlike signal along the worldsheet is able to reach could be complicated, and we will not try to figure out what it is, except to restate that it should be bounded by the lightlike spacetime geodesic studied in section~\ref{SPACETIME}.  A lightlike geodesic on the string worldsheet, not in bulk spacetime, is relevant for finding out which parts of the trailing string remain undeformed at a given time $t>0$ because particles like gravitons traveling on lightlike spacetime geodesics can affect the string only through string interactions, which are suppressed by $1/N$.

We already calculated the tangent vectors to lightlike geodesics in \eno{lkDef}.  The relevant one is $\ell^\alpha$ because it goes down the string worldsheet (or goes down ``faster'' in the region where $h<v^2$).  The differential equation $dt/dz = \ell^t$ is readily integrated, and it gives
 \eqn{tOfz}{
  {t \over z_H} = {1 \over 4} \log {1+y \over 1-y} +
    {i \over 4} \log {1+iy \over 1-iy} -
    {i\sqrt\gamma \over 2} \log {1+iy\sqrt\gamma \over
      1-iy\sqrt\gamma} \,,
 }
where $\gamma = 1/\sqrt{1-v^2}$ as usual.  Plugging \eno{tOfz} into \eno{Xembed}, one finds the orbit $x^1 = x^1(z)$ in spacetime of the point below which the trailing string must be undisturbed:
 \eqn{POND}{
  \hat{x}^1_{\rm POND} = {x^1 \over z_H} =
    {v \over 2i} \sqrt\gamma \log
      {1+iy\sqrt\gamma \over 1-iy\sqrt\gamma} \,.
 }
POND stands for point of no disturbance.  $\Delta\hat{x}_{\rm POND}$ is the difference between $\hat{x}^1_{\rm POND}$ when $y=1$ and its initial value when $y=y_{\rm UV}$.

Whatever the string may do above the point of no disturbance, $\Delta\hat{x}_{\rm POND}$ sets a lower bound on how far it gets in the $x^1$ direction.  As is evident from figure~\ref{DeltaXProfilePlot}, $\Delta\hat{x}_{\rm POND}$ is almost as big as $\Delta\hat{x}_{\rm spacetime}$ for $v$ close to $1$.  And because $\Delta\hat{x}_{\rm POND}$ is a function of $v$ and $y_{\rm UV}$, it can be passed through the same maximization procedure as described at the end of section~\ref{SPACETIME} to give estimates $\Delta \hat{x}_A(\hat{E})$ of the maximum penetration depth, where $A$ includes ``POND.''  A caveat is that $\Delta\hat{x}_{\rm POND}$ involves an evaluation of the point of no disturbance in the limit $y \to 1$, where it crosses the horizon; but the estimate $E_{\rm trailing}$ of the energy of the string requires an infrared cutoff at some finite value $y_{\rm IR} < 1$.  Evidently, there is an inconsistency in whether or not we include in the calculations the part of the trailing string closer to the horizon than $y_{\rm IR}$.  This caveat can be avoided by using the fixed $x^1$ strategy for evaluating the initial energy.  But to check that it is unlikely to influence our qualitative conclusions when we use an infrared regulator to calculate the energy, we considered an alternative definition of $\Delta\hat{x}_{\rm POND}$ which is the difference of $\hat{x}^1_{\rm POND}$ evaluated at $y_{\rm IR}$ and $y_{\rm UV}$ rather than at $1$ and $y_{\rm UV}$.  This is indicated by including ``cutoff'' in $A$.

Figure~\ref{CombinedPlot} includes numerical evaluations of $\Delta \hat{x}_A$ for two different sets of assumptions including POND.  We again see that the functional dependence $\Delta \hat{x}_A(\hat{E})$ is roughly the same for the different assumption sets.  For $A = \{{\rm POND},\textrm{fixed $x^1$}\}$, analytic approximations at large $\hat{E}$ give
 \eqn{LargeEWorldsheet}{
  \Delta \hat{x}_{{\rm POND},\textrm{fixed $x^1$}} &= 0.8798 \hat{E}^{1/3} - 0.8252 + 0.058 \hat{E}^{-1/3} + 0.582 \hat E^{-2/3} - 1.601 \hat E^{-1} \cr
   {}&\qquad{} + 2.25 \hat E^{-4/3} - 1.16 \hat E^{-5/3} - 2.11 \hat E^{-2} + {\cal O}(\hat E^{-7/3})
 }
To derive \eno{LargeEWorldsheet}, the first step is to write the relevant energy estimate, \eno{Efixedx1}, as
 \eqn{Efixedx1Again}{
  \hat{E} = \gamma \left( {1 \over y_{\rm UV}} - 1 \right) \,.
 }
Consider now the combination
 \eqn{lambdaDef}{
  \lambda \equiv \gamma y_{\rm UV}^2 \,.
 }
The relations \eno{Efixedx1Again} and \eno{lambdaDef} may be inverted to express $\gamma$ and $y_{\rm UV}$ in terms of $\hat{E}$ and $\lambda$.  The next step is to use \eno{POND} to express $\Delta\hat{x}_{\rm POND}$ in terms of $\hat{E}$ and $\lambda$. This can
be done in closed form, but the explicit expression is not very illuminating. The maximum of $\Delta\hat{x}_{\rm spacetime}$ is attained at a value $\lambda=\lambda_*$ determined by the equation
 \eqn{ExtremeEquation}{
  \left( {\partial \over \partial\lambda}
    \Delta x_{\rm spacetime}(\hat{E},\lambda)
    \right)_{\lambda=\lambda_*} = 0 \,.
 }
The asymptotic behavior of $\lambda_*$ can be calculated by plugging a large $\hat{E}$ expansion of the form
 \eqn{lambdastarLargeE}{
  \lambda_* = \lambda_*^{(0)} + \lambda_*^{(1)} \hat E^{-1/3} + \lambda_*^{(2)} \hat E^{-2/3} + \cdots
 }
into \eno{ExtremeEquation}and solving for the coefficients $\lambda_*^{(i)}$ term by term. For example,
setting the coefficients of the first two terms to zero leads to the equations
 \eqn{TwoLambdas}{
 -3 \sqrt{\lambda_*^{(0)}} +
 \left( 1+\lambda_*^{(0)} \right) \cot^{-1} \sqrt{\lambda_*^{(0)}} &= 0 \cr
 2 \left( \lambda_*^{(0)} \right)^{4/3} \left( 1+ \lambda_*^{(0)} \right) +
 \left(-2+\lambda_*^{(0)}\right) \lambda_*^{(1)} &=0 \,.
}
These equations can be solved numerically to give $\lambda_*^{(0)}=0.212$ and $\lambda_*^{(1)}=0.171$.  Finally, plugging the series expansion \eno{lambdastarLargeE} back into our expression for
$\Delta\hat{x}_{\rm POND}(\hat E,\lambda)$ and expanding at large $\hat{E}$, one recovers \eno{LargeEWorldsheet}.

The method we just described can also be used to obtain the large $\hat E$ behavior of $\Delta\hat{x}_{\rm spacetime}$ given in \eno{LargeESpacetime}.

\section{Comparing to BDMPS estimates of energy loss}
\label{COMPARISONS}

A standard way of estimating energy loss by hard partons is the BDMPS jet-quenching formalism, which is summarized for example in \cite{Liu:2006he,CasalderreySolana:2007pr}.  The energy loss of a hard parton in a representation $R$ of the color group $SU(N)$ is
 \eqn{DeltaEBDMPS}{
  \Delta E = {1 \over 4} \alpha_s C_R \hat{q} (\Delta x)^2 \,,
 }
where $\Delta x$ is the distance traveled.\footnote{Some authors prefer to measure distance traveled using a lightcone coordinate, which introduces an additional factor of $1/2$ into \eno{DeltaEBDMPS}.}  $C_R$ is the Casimir denoted $C_2(R)$ on p.~500ff of \cite{Peskin:1995ev}, so that $C_F = (N^2-1)/2N$ for a fundamental quark and $C_A = N$ for a gluon.  A perturbative estimate gives
 \eqn{qhatPerturbative}{
  \hat{q}_{\rm pert} = {8\zeta(3) \over \pi} \alpha_s^2 N^2 T^3 \,.
 }
A rule-of-thumb estimate for the strong coupling is $\alpha_s = 1/2$.

A calculation in ${\cal N}=4$ SYM starting from a Wilson loop definition of $\hat{q}$ yields \cite{Liu:2006ug}
 \eqn{qhatLRW}{
  \hat{q}_{\rm LRW} = {\pi^{3/2} \Gamma(3/4) \over \Gamma(5/4)}
    \sqrt{g_{YM}^2 N} \, T^3 \,.
 }
It seems clear that QCD should exhibit a lower value of $\hat{q}$ because it has fewer degrees of freedom: about a third as many as measured by the entropy density.  One way to incorporating this factor was proposed in \cite{Liu:2006he}: it is to include a proportionality to the square root of the entropy in $\hat{q}$.  Thus
 \eqn{qhatLRWestimate}{
  \hat{q}_{\rm scaled\ LRW} \approx \sqrt{47.5\over 120}\, \hat{q}_{\rm LRW} \approx
    0.63 \, \hat{q}_{\rm LRW} \,.
 }
Because $\hat{q}_{\rm scaled}$ still depends on the 't~Hooft coupling $g_{YM}^2 N$, one must fix the value of this coupling.  An obvious way to do so is to insist that tree-level gluon scattering processes should have the same amplitude in ${\cal N}=4$ gauge theory as in QCD: that is, the tree-level couplings coincide, resulting in $g_{YM}^2 N = 6\pi$ when $\alpha_s = 1/2$ and $N=3$.  In quoting a numerical value for $\hat{q}_{\rm scaled\ LRW}$ in table~\ref{QhatSummary}, we have used the ``obvious scheme:'' $g_{YM}^2 N = 6\pi$ and $T_{{\cal N}=4}=T_{QCD}$.

An alternative scheme for comparing ${\cal N}=4$ SYM to QCD was proposed in \cite{Gubser:2006qh}.  Instead of comparing at fixed temperature, one compares at fixed energy density.  This is supposed to correct, approximately, for the larger number of degrees of freedom in ${\cal N}=4$ theory, and it approximately amounts to setting $T_{{\cal N}=4} = T_{\rm QCD}/3^{1/4}$.  Thus, in place of \eno{qhatLRWestimate}, one would have
 \eqn{qhatLRWalternative}{
  \hat{q}_{\rm alternative\ LRW} \approx
   {1 \over 3^{3/4}} \hat{q}_{\rm LRW} \,.
 }
Also, instead of comparing at fixed tree-level coupling, one chooses in the alternative scheme a value $g_{YM}^2 N \approx 5.5$ in order to approximately match the force between a heavy quark and anti-quark separated by a distance on order $0.25\,{\rm fm}$ in a medium at a temperature characteristic of RHIC collisions.  (Even smaller values of $g_{YM}^2 N$ can be motivated by comparing SYM and QCD at fixed Debye mass \cite{Bak:2007fk}.)  As shown in table~\ref{QhatSummary}, $\hat{q}_{\rm alternative\ LRW}$, with $g_{YM}^2 N = 5.5$, is essentially indistinguishable from the perturbative estimate \eno{qhatPerturbative}.

To compare the BDMPS result to the falling string calculations, we use the operational definition \eno{qhatOperational}, which amounts to setting the stopping length of a hard gluon equal to the value of $\Delta x$ that one obtains from \eno{DeltaEBDMPS} upon setting $\Delta E$ equal to the initial energy of the gluon (and, of course, $C_R=3$).  We further set $\alpha_s = 1/2$ in \eno{qhatOperational}, obtaining
 \eqn{qhatRough}{
  \hat{q}_{\rm rough} = {8E \over 3 (\Delta x)^2} \,.
 }
For $x$ in \eno{qhatRough}, we plug in a value estimated from string theory in one of the ways we have explained above.  Using also \eno{HattedDefs}, we find that $\hat{q}_{\rm rough}$ becomes
 \eqn{qhatFromFall}{
  \hat{q}_{\rm fall} = {8\pi^2 \over 3} \sqrt{g_{YM}^2 N}
    {\hat{E} \over \Delta\hat{x}(\hat{E})^2} T^3 \,.
 }
We emphatically warn the reader that \eno{qhatFromFall} is only a rough estimate of the ``effective $\hat{q}$'' implied by our falling string picture, because the underlying physical picture is significantly different from the BDMPS formalism.  Let us review the differences before proceeding to extract numbers from \eno{qhatFromFall}:
 \begin{enumerate}
  \item The ``gluon'' as described by the falling string is off-shell: it follows a time-like trajectory.  This contrasts with the eikonal approximation of lightlike trajectories employed in the BDMPS treatment.
  \item In the zero-temperature calculation of \cite{Alday:2007hr}, the string worldsheet can be understood to arise, in the usual sense of 't~Hooft, from a sum over an infinite set of planar diagrams contributing to a certain exclusive process; similarly, in our treatment, the strong interactions of the gluon with the medium are encoded in the classical dynamics of the worldsheet.  This again contrasts with BDMPS, which is a partially perturbative treatment of radiative energy loss to the medium.  It is not obvious to us how to translate some aspect of the falling string calculation to the spectrum of radiated energy.
  \item We have not included fluctuations in our treatment, so we cannot (yet) give an account of the diffusion of transverse momentum similar to the one that is a prominent part of the BDMPS formalism.
 \end{enumerate}

\begin{figure}
  \centerline{\includegraphics[width=6.5in]{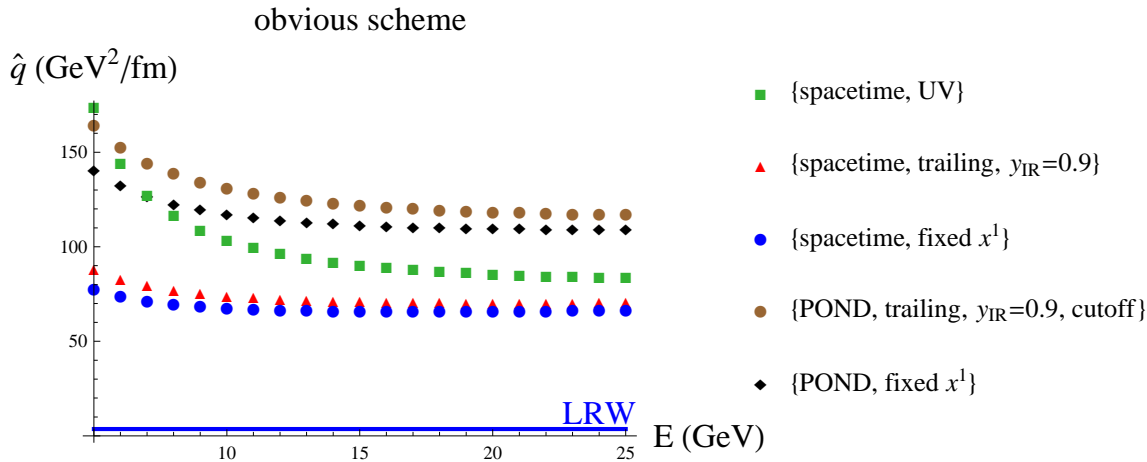}}
  \caption{Evaluations of ${\hat q}$ as given in \eqref{qhatFromFall} under a variety of assumptions, using the obvious scheme.  The solid blue curve corresponds to the LRW prediction \eqref{qhatLRW}.}\label{ObviousPlot}
 \end{figure}

\begin{figure}
  \centerline{\includegraphics[width=6.5in]{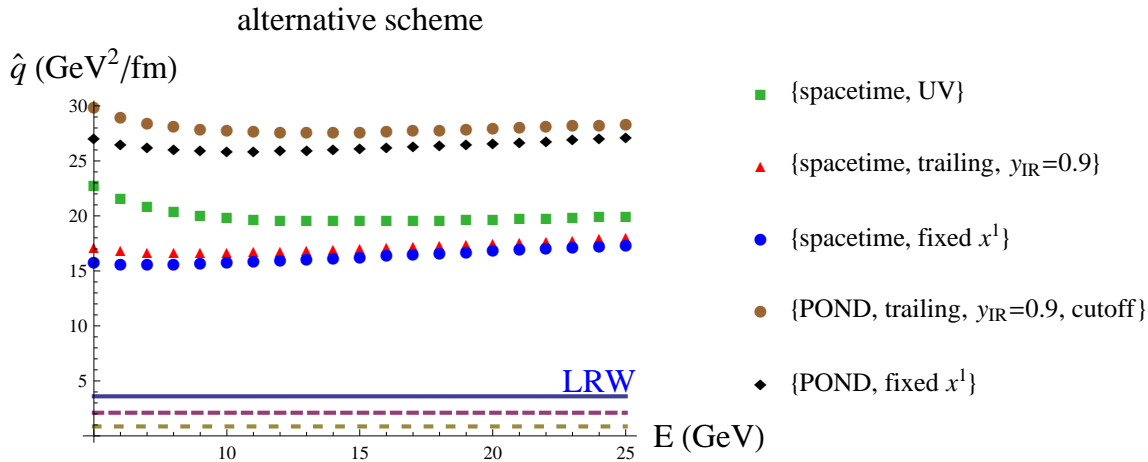}}
  \caption{Evaluations of ${\hat q}$ as given in \eqref{qhatFromFall} under a variety of assumptions, using the alternative scheme.  The solid blue curve corresponds to the LRW prediction \eqref{qhatLRW}, the dashed purple curve corresponds to the ``scaled LRW'' prediction \eqref{qhatLRWestimate}, and the dotted yellow curve corresponds to the ``alternative LRW'' estimate \eqref{qhatLRWalternative}.}\label{AlternativePlot}
 \end{figure}

Either the obvious scheme ($T_{QCD}=T_{{\cal N}=4}$ and $g_{YM}^2 N = 6\pi$) or the alternative scheme ($T_{QCD}=T_{{\cal N}=4}/3^{1/4}$ and $g_{YM}^2 N = 5.5$) can be applied to \eno{qhatFromFall}, and we will denote the resulting expressions $\hat{q}_{\rm obvious\ fall}$ and $\hat{q}_{\rm alternative\ fall}$.  In both schemes we take $T_{QCD} = 280\,{\rm MeV}$, which is a reasonable estimate for central gold-gold collisions at RHIC's top energy, $\sqrt{s}_{NN} = 200\,{\rm GeV}$.  Because ${\hat x}$ is not exactly proportional to $\hat E^{1/2}$, ${\hat q}$ in equation \eqref{qhatFromFall} depends on ${\hat E}$.  However, this dependence is rather weak, amounting at large enough $\hat{E}$ to a $\hat{E}^{1/6}$ behavior.  Figures~\ref{ObviousPlot} and~\ref{AlternativePlot} show that for \mbox{$5\,{\rm GeV} \leq E \leq 25\,{\rm GeV}$}, which we take as a representative range of energies for hard gluons in the QGP produced at RHIC, $\hat{q}$ is roughly constant.
By fitting $\hat x$ to square-root dependence on $\hat E$ for $E$ in the above-mentioned range, and using \eqref{qhatFromFall}, we find
 \eqn{qhatBottomLine}{\seqalign{\span\TL & \span\TR &\qquad
   \span\TL & \span\TR}{
  \hat{q}^{\{\textrm{POND, fixed $x^1$}\}}_\textrm{obvious fall}
    &= 116\,{{\rm GeV}^2 \over {\rm fm}} &
  \hat{q}^{\{\textrm{spacetime, fixed $x^1$}\}}_\textrm{obvious fall}
    &= 68\,{{\rm GeV}^2 \over {\rm fm}}  \cr
  \hat{q}^{\{\textrm{POND, fixed $x^1$}\}}_\textrm{alternative fall}
    &= 26\,{{\rm GeV}^2 \over {\rm fm}} &
  \hat{q}^{\{\textrm{spacetime, fixed $x^1$}\}}_\textrm{alternative fall}
    &= 16\,{{\rm GeV}^2 \over {\rm fm}}
 }}
We chose the fixed $x^1$ computation of energy, \eno{Efixedx1}, because it avoids the use of IR cutoff and thus incorporates fewer assumptions.

Representative numerical values for the estimates of $\hat{q}$ discussed here are presented in table~\ref{QhatSummary}.  The values of $\hat q_{\rm obvious\ fall}$ and $\hat q_{\rm alternative\ fall}$ that appear in this table are the averages of the corresponding quantities calculated from POND and spacetime geodesics, as quoted in \eno{qhatBottomLine}.  We again stress that these averages should be regarded as rough estimates.
 \begin{table}
 \begin{center}
 \begin{tabular}{c||c|c|c|c|c|c}
  quantity & $\hat{q}_{\rm pert}$ & $\hat{q}_{\rm LRW}$ &
    $\hat{q}_{\rm scaled\ LRW}$ & $\hat{q}_{\rm alternative\ LRW}$ &
    $\hat{q}_{\rm alternative\ fall}$ & $\hat{q}_{\rm obvious\ fall}$
     \\ \hline
  value @ $280\,{\rm MeV}$ &
   0.77 & 3.6 & 2.3 & 0.86 & 21 & 92
 \end{tabular}
 \end{center}
 \caption{Various ways of estimating $\hat{q}$.  See the main text for details in each one.  All values for $\hat{q}$ are quoted in units of ${\rm GeV}^2/{\rm fm}$.}\label{QhatSummary}
 \end{table}

As remarked around \eno{PHENIXqhat}, a recent comparison of parton quenching models to PHENIX data favors the range $\hat{q} \approx 7-28\,{\rm GeV}^2/{\rm fm}$ \cite{Adare:2008cg}.  It is interesting, but in no way conclusive, that the only estimate of $\hat{q}$ in table~\ref{QhatSummary} that falls within this range is $\hat{q}_{\rm alternative\ fall}$.  Of the estimates based on \cite{Liu:2006ug}, it is fair to exclude $\hat{q}_{\rm LRW}$ because it is intended to be for ${\cal N}=4$ SYM, without any rescalings that would account for the differences between ${\cal N}=4$ and QCD.  So---with $T=280\,{\rm MeV}$---both perturbative estimates and those based on \cite{Liu:2006ug} come out below the $3\sigma$ range \eno{PHENIXqhat}, and $\hat{q}_{\rm obvious\ fall}$ comes out above.  But the $T^3$ dependence of $\hat{q}$ makes it difficult to pin down the theoretical predictions with much precision.

We are pleased to see a certain consistency emerging between the falling string calculations of this paper and the computations of heavy quark drag and diffusion in \cite{Herzog:2006gh,Casalderrey-Solana:2006rq,Gubser:2006bz}: whereas the energy predicted by string theory is too strong when compared in the obvious scheme, it is close---though perhaps still a bit high---when compared in the alternative scheme \cite{Gubser:2006qh}.  Other comparison analyses have been proposed in the heavy quark setup which give pretty good agreement between string predictions and data: see for example \cite{Chesler:2006gr}, in which perturbative results are compared between ${\cal N}=4$ SYM and QCD as well as strong coupling results.

The BDMPS approach to energy loss is hardly the only one in common use.  Others include the higher twist, GLV, and AMY formalisms \cite{Wang:2003mm,Gyulassy:2000er,Arnold:2002ja,Bass:2008rv}.  Each approach makes a different set of assumptions.  It would be useful to make cross-comparisons with the falling string.  We leave this task for future work.

\section{The null string limit}
\label{NULL}

They key to obtaining an estimate for the stopping length $\Delta\hat{x}(\hat{E})$ is that the POND trajectory, which provides an approximate lower bound on $\Delta\hat{x}$, is close to the massless spacetime geodesic, which provides an approximate upper bound on $\Delta\hat{x}$: see figure~\ref{FallingString}.  To see why this happens, consider the limit $v \to 1$ with $y_{\rm UV}$ held fixed.  A straightforward calculation shows that the POND trajectory coincides with the spacetime geodesic in this limit.  The key point is that the worldsheet becomes a null surface in this limit, so the lightlike tangent vector $\ell^\mu$ to the POND trajectory is also normal to the worldsheet.  In fact, in this limit, $\ell^\mu$, $k^\mu$, and $n^\mu$ all coincide up to overall magnitudes.  A heuristic way of thinking about this is that in the $v \to 1$, the string is replaced by an ensemble of massless particles, all following critical trajectories of the form \eno{SpecialOrbit} (but with different values of $K$).  Signals can't propagate up or down the string in this limit: every ``bit'' of string is causally isolated from every other bit, and follows a massless spacetime geodesic.  We can develop this heuristic picture by defining a scaled version of the string's inverse tension:
 \eqn{alphaScaled}{
  \alpha'_{\rm scaled} \equiv \alpha' \sqrt{1-v^2} \,,
 }
Then, formally, we can take $v \to 1$ and $\alpha' \to \infty$ in such a way that $\alpha'_{\rm scaled}$ remains fixed.  We will describe this as the ``null string'' limit, because the string worldsheet becomes a null surface.  It is a formal limit because when $\alpha' \gg L^2$, stringy corrections to supergravity probably become large, so the $AdS_5$-Schwarzschild background is expected to be significantly altered.  But it captures the key idea that the string tension doesn't matter in the $v \to 1$ limit.

As an application of the null string limit, we can compute the five-dimensional stress tensor of the falling string for $v \to 1$.  We are ignoring stringy corrections, so the action we start with is
 \eqn{GravityPlusMatter}{
  S = {1\over 2 \kappa_5^2} \int d^5 x \, \sqrt{-G} \left[ R + {12 \over L^2} \right] +
    S_M \,,
 }
where $S_M$ is the matter action.  The five-dimensional stress tensor $\tau^{\mu\nu}$ can be defined through the equation
 \eqn{DefineStress}{
  \delta S_M = \int d^5 x \, \sqrt{-G} \, \delta G_{\mu\nu}
    {1 \over 2} \tau^{\mu\nu} \,,
 }
so that it enters into the Einstein equation as
 \eqn{EinsteinEq}{
  R^{\mu\nu} - {1 \over 2} R G^{\mu\nu} - {6 \over L^2} G^{\mu\nu} =
    \kappa_5^2 \tau^{\mu\nu} \,.
 }
The result will be that the null string's stress tensor is an integral of the stress tensors for continuously many massless particles propagating on critical null geodesics.

First, let's compute the stress tensor of a massless particle, starting from the action \eno{AffineAction} with $m=0$.  The result is immediate:
 \eqn{ParticleStress}{
  \tau_{\rm particle}^{\mu\nu} &= {1 \over \sqrt{-G}} \int d\eta \,
    \delta^5(x^\mu - X^\mu(\eta)) {1 \over e} {dX^\mu \over d\eta}
      {dX^\nu \over d\eta}  \cr
   &= {1 \over \sqrt{-G}} \int d\eta \, \delta^5(x^\mu-X^\mu(\eta))
     {dX^\mu \over d\eta} p^\nu  \cr
   &= {1 \over \sqrt{-G}} \delta^4(x^m-X^m(z)) {dX^\mu \over dz}
    p^\nu \,,
 }
where $p_\mu = {1 \over e} G_{\mu\nu} {dX^\nu \over d\eta}$ is the momentum conjugate to $X^\mu$, and in the last line we have specialized to the gauge $\eta=z$.  The quantities $p_0$ and $p_1$ coincide with the expressions in \eno{Momenta} when $\eta=z$.  Let us now specialize to the critical trajectory \eno{SpecialOrbit} and set $K = {x^1_H \over z_H} + 1$, so that $x^1_H$ is the position at which the geodesic crosses the horizon.  Introducing the one-form
 \eqn{TangentToGeodesic}{
  b_\mu = \begin{pmatrix} -1 & 1 & 0 & 0 & {\sqrt{1-h} \over h}
    \end{pmatrix} \,,
 }
which is tangent to the particle's trajectory, and employing the gauge $\eta=z$, it is straightforward to show that
 \eqn{TauLower}{
  \tau^{\mu\nu}_{\rm particle} =
    \tau_{\rm particle}(x^\mu;x^1_H) b^\mu b^\nu
 }
where
 \eqn{tauExpress}{
  \tau_{\rm particle}(x^\mu;&x^1_H) = {E z z_H^2 \over L^3}
     \delta\left( t-x^1_H - z_H + {z_H^2 \over z} +
     \xi(z) \right)
    \delta\left( x^1-x^1_H - z_H + {z_H^2 \over z} \right)
    \delta(x^2) \delta(x^3) \,.
 }
The product of delta functions in \eno{tauExpress} is simply $\delta^4(x^m-X^m(z))$, and we set $E = -p_0$.

Next, we can compute the stress tensor of a string with $v<1$ starting from the Nambu-Goto action \eno{SNG}.  The result is
 \eqn{tauString}{
  \tau^{\mu\nu}_{\rm string} = {1 \over \sqrt{-G}} \int d^2 \sigma \,
    \sqrt{-g} \, \delta^5(x^\mu - X^\mu(\sigma)) \, \partial_\alpha X^\mu
     P^{\alpha\nu}
 }
where $P_\mu^\alpha$ is given by \eno{Palphamu}.  Taking the null string limit, one finds
 \eqn{StringP}{
  p_\mu^\alpha \equiv
    \lim_{\rm null\atop string} \sqrt{-g} P^\alpha_\mu =
    {L^2 \over 2\pi \alpha'_{\rm scaled}} {1 \over z_H^2 h}
    \begin{pmatrix} -{1 \over \sqrt{1-h}} &
      {1 \over \sqrt{1-h}} & 0 & 0 & {1 \over h} \\
     -h & h & 0 & 0 & \sqrt{1-h} \end{pmatrix} \,.
 }
Both rows of \eno{StringP} are proportional to $b_\mu$ as defined in \eno{TangentToGeodesic}, so the matrix has rank 1.  Observing that $b_\mu$ coincides, up to an overall factor, with $n_\mu$, $\ell_\mu$, and $k_\mu$ in the limit $v \to 1$, it already seems inevitable that the stress tensor of the null string will reduce to an ensemble of massless particles.  Indeed, by plugging \eno{StringP} into \eno{tauString} one finds
 \eqn{StressInt}{
  \tau^{\mu\nu}_{\rm null\atop string} \equiv
   \lim_{\rm null\atop string} \tau^{\mu\nu}_{\rm string} =
   {1 \over \sqrt{-G}}
   \theta\left( x^1_H - x^1 + z_H - {z_H^2 \over z} \right)
    \delta(t - x^1 + \xi(z)) \delta(x^2) \delta(x^3)
    \partial_\alpha X^\mu p^{\alpha\nu}
 }
where
 \eqn{thetaDef}{
  \theta(x) = \left\{ \seqalign{\span\TL &\qquad\span\TT}{
   0 & for $x < 0$  \cr  1 & for $x>0\,.$} \right.
 }
The first delta function in \eno{StressInt} enforces the defining relation \eno{Xembed} of the $v \to 1$ limit of the trailing string.  The theta function factor arises because the boundary of the null string follows the orbit \eno{SpecialOrbit} with $K = {x_H^1 \over z_H} + 1$.  Using \eno{StringP} and \eno{TrailingEnergy},
 \eqn{NiceStress}{
  \tau^{\mu\nu}_{\rm null\atop string} =
    \tau_{\rm string}(x^\mu;x^1_H) b^\mu b^\nu \,,
 }
where
 \eqn{NiceTau}{
  \tau_{\rm string}(x^\mu;x^1_H) =
    {z/L \over 2\pi\alpha'_{\rm scaled}}
   \theta\left( x^1_H - x^1 + z_H - {z_H^2 \over z} \right)
    \delta(t - x^1 + \xi(z)) \delta(x^2) \delta(x^3) \,.
 }
To see that the stress tensor of the string is identical to the stress tensor of an ensemble of massless particles following critical trajectories, we need only note that
 \eqn{DerivativeRelation}{
  {\partial \tau_{\rm string}(x^\mu;x^1_H) \over
    \partial x^1_H} = {1 \over 2\pi\alpha'_{\rm scaled}}
      {L^2 \over E z_H^2} \tau_{\rm particle}(x^\mu;x^1_H) \,.
 }
Turning \eno{DerivativeRelation} around, $\tau_{\rm string}$ is an integral over $x^1_H$ of $\tau_{\rm particle}$, and since $b^\mu b^\nu$ has no explicit dependence on $x^1_H$, the same relation holds between $\tau^{\mu\nu}_{\rm string}$ and $\tau^{\mu\nu}_{\rm particle}$.

A consequence of the discussion of the last few paragraphs is that to compute the contribution of a falling null string to the expectation $\langle T_{mn} \rangle$ of the gauge-theory's stress tensor, one can start by doing the analogous calculation for a massless particle and then integrate with respect to $x^1_H$.  The falling null string is, as we have discussed, only a formal approximation to the finite-tension falling strings of real physical interest.  But since its shape is known analytically, it seems a worthwhile starting point for an investigation of $\langle T_{mn} \rangle$.  We hope to report on calculations along these lines in future work.

So far in this section, we have focused on the limit $v \to 1$ with $y_{\rm UV}$ held fixed.  But in studying the maximum penetration length of high-energy probes, a different limit is appropriate: for fixed energy, we maximize $\Delta x$ by varying $v$ and $y_{\rm UV}$ so as to hold $E$ fixed.  For large $\hat{E}$, the maximum is attained for $\gamma y_{\rm UV}^2 = \lambda_* \approx \lambda_*^{(0)}$, where $\lambda_*^{(0)} = 0.154$ for $A = \{{\rm spacetime}, {\rm fixed}\ x^1\}$ and $0.212$ for $A = \{{\rm POND}, {\rm fixed}\ x^1\}$.  Thus, the limit of interest for estimating penetration length of very hard probes is $v \to 1$ with $\gamma y_{\rm UV}^2$ held fixed---but held fixed to different values for spacetime as compared to POND.  In this $\hat{E} \to \infty$, the maximum $\Delta x_{\rm POND}$ and the maximum $\Delta x_{\rm spacetime}$ do not approach one another: instead, $\Delta x_{\rm POND} / \Delta x_{\rm spacetime} \to 0.86$.

\section{Discussion}
\label{DISCUSSION}

The trailing string of \cite{Herzog:2006gh,Gubser:2006bz} is essentially an equilibrium configuration, where energy is lost at a constant rate into the plasma, but the quark never slows down because it is infinitely massive.  A key feature of heavy-ion physics is that many or even most hard partons travel only a short distance through the medium before substantially stopping.  Our discussion of falling strings is a first attempt to incorporate into the trailing string picture the dramatically non-equilibrium nature of the physics of energy loss for light partons.  It may help the reader's intuition to note that in the alternative scheme, $\hat{x}=2$ corresponds to $x \approx 0.6\,{\rm fm}$, and $\hat{E}=20$ corresponds to $E \approx 10\,{\rm GeV}$, when $T=280\,{\rm MeV}$ in the real-world plasma.  So, according to figure~\ref{CombinedPlot}, a $10\,{\rm GeV}$ gluon stops in a distance of about $0.5\,{\rm fm}$.

Finally, we can consider how the falling string picture might generalize to theories with fundamental quarks whose mass is finite.  In the construction of \cite{Karch:2002sh}, fundamentally charged quarks come from strings stretched between the D3-branes and D7-branes, where the D3-branes create the $AdS_5$ (or $AdS_5$-Schwarzschild) geometry, and the D7-branes are usually treated in the probe approximation.  At finite temperature, the D7-branes either descend to a minimum distance from the black hole horizon if the mass of the corresponding quark is sufficiently larger than the temperature, or they extend down into the horizon if the corresponding quark is light \cite{Mateos:2006nu}.  (In any case, provided temperature is constant, the D7-branes are static, and apparently stable.)  At the risk of oversimplifying, let's ignore the geometry of the D7-brane embeddings in ten dimensions and replace them by ``flavor-branes'' that fill $AdS_5$-Schwarzschild either down to some maximal depth $z_* < z_H$, or that extend across the horizon at $z=z_H$.  The finite-mass trailing string as considered in \cite{Herzog:2006gh} was assumed to end on a flavor brane at the maximum possible depth, i.e.~$z=z_*$.  A curious property of such strings is that they have a maximum speed, $v = \sqrt{1-z_*^4/z_H^4}$, which we alluded to following \eno{FoundPalphamu}.  This speed gets smaller as the quark mass gets smaller, and it is in some sense zero for quarks that are light enough so that the corresponding D7-brane extends into the horizon.  It now seems clear that this maximum speed should be understood as the speed above which a falling string picture of energy loss must be taken into account.  More precisely, an energetic quark can be represented as a string coming out of the horizon and extending up to a lesser depth, $z_{\rm UV} < z_*$, than the maximum depth of the flavor brane (if there is one).  Essentially the same analysis we have given for gluons could be replayed for such strings, with the main difference being that the string is no longer doubled, but ends on the flavor brane.  Admittedly, a fully correct treatment of the D3-D7 construction would involve non-trivial motion of the string in the full ten-dimensional geometry, such that the projection to $AdS_5$-Schwarzschild would {\it not} have the property that the string endpoint travels on a null trajectory.  However, we are inclined to think that such a motion, which involves the $SO(6)$ R-symmetry of ${\cal N}=4$ super-Yang-Mills theory, doesn't translate very precisely to QCD.  It may be that ignoring the ten-dimensional geometry altogether and employing flavor-branes in place of bona fide string theory constructions captures approximately the right physics.  For light quarks, where there is no maximum depth, the only change would then be to replace $E$ by $E/2$: that is, energy loss is half as fast for fundamentally charged light quarks as for gluons.  This is the same scaling as found in a BDMPS treatment, where according to \eno{DeltaEBDMPS} the energy scales linearly with $C_R$, and
 \eqn{CFCA}{
  {C_F \over C_A} = {N^2-1 \over 2N^2} \approx {1 \over 2} \,,
 }
where $N=3$ is the number of colors.  For heavy quarks, there could be a two-stage process of energy loss, where one first has $\Delta E \propto (\Delta x)^3$ due to falling string dynamics, and then, after the string has fallen as far as it can while still remaining attached to the flavor brane, one has $dE/dx \propto p$, characteristic of trailing string dynamics.

\begin{figure}
  \centerline{\includegraphics[width=6in]{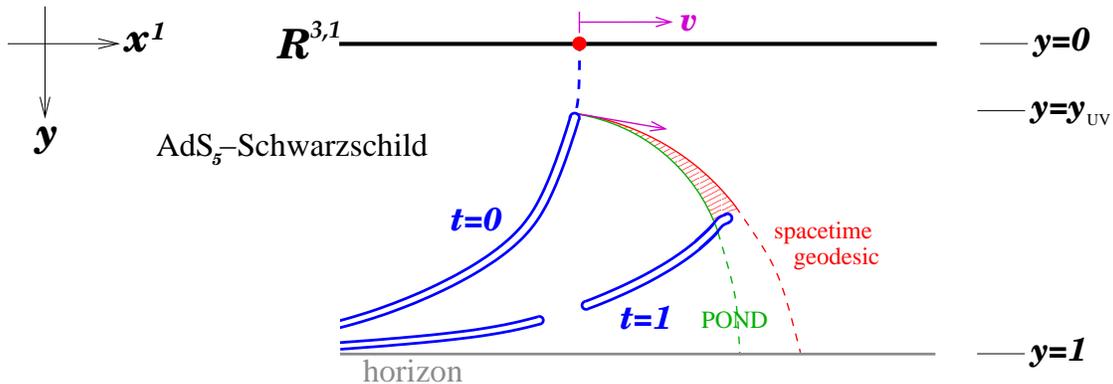}}
  \caption{A doubled falling string which experiences a splitting interaction at a time between $t=0$ and $t=1$.  This interaction is suppressed by one power of $N$.}\label{CutString}
 \end{figure}

We have assumed throughout that string splitting and string joining interactions are negligible.  In fact they are suppressed by a power of $N$.  In the context of doubled strings, the string could split anywhere along its length, and the amplitude to do so is proportional to $g_{\rm str} \propto g_{YM}^2$, which is indeed an ${\cal O}(1/N)$ effect using 't~Hooft scaling.  However, this suppression may not be enough to make the effect unimportant.  The result of one such splitting is illustrated in figure~\ref{CutString}.  Physically, the splitting describes a decoupling of most of the hard parton's momentum from its color---a sort of dual hadronization in the fifth dimension, where the closed string that carries most of the momentum is a color-singlet glueball in the process of thermalizing.  If splitting is significant, it might help the string get a little further in the $x^1$ direction, but no further than the spacetime geodesic considered in section~\ref{SPACETIME}.  Splitting becomes less easy if the string is not perfectly doubled, as one must expect in a more realistic treatment.

We should keep in mind that falling strings as we have considered them in this paper may be considerably idealized in comparison with strings that form as a result of energetic collisions.  The latter are probably less orderly, and they may be less optimized to travel long distances before disappearing into the horizon.  This might push our estimates of $\hat{q}$ upward.  On the other hand, we have not accounted for fluctuations, nor have we calculated $\langle T_{mn} \rangle$.  Either of these elaborations might be phenomenologically significant, and the latter might be affected by the string splitting interactions discussed above.

\section*{Acknowledgments}

We thank M.~Gyulassy, A.~Majumder, T.~Tesileanu, A.~Yarom, and B.~Zajc for useful discussions and the nuclear physics group at Columbia University for questions and comments on a preliminary version of the results in this paper.  This work was supported in part by the Department of Energy under Grant No.\ DE-FG02-91ER40671 and by the NSF under award number PHY-0652782.  F.D.R.~was also supported in part by the FCT grant SFRH/BD/30374/2006.  The work of D.R.G.~was also supported in part by the NSF Graduate Research Fellowship Program.

\bibliographystyle{ssg}
\bibliography{fall}
\end{document}